\documentclass[11pt]{article}
\usepackage{amssymb}
\usepackage{amsmath}
\usepackage{color}
\usepackage[nocompress]{cite}

\begin{document}

\title{\hfill {\small \medskip }\\
\textbf{Constant curvature black holes in Einstein AdS gravity: conserved
quantities.}}
\author{Pablo Guilleminot$^{a}$, Rodrigo Olea$^{a}$ and Alexander N. Petrov$^{b}$\smallskip \\
%EndAName
$^{a}${\small \emph{Departamento de Ciencias F\'{\i}sicas, Universidad
Andres Bello,}}\\
{\small \emph{Sazi\'e 2212, Piso 7, Santiago, Chile. \smallskip }}\\
$^{b}$\emph{\small Moscow MV Lomonosov State University, Sternberg
Astronomical Institute, }\\
\emph{\small Universitetskii pr., 13, Moscow 119992, Russia.}{\small \emph{%
\smallskip }}\\
{\small \texttt{yemheno@gmail.com, rodrigo.olea@unab.cl, alex.petrov55@gmail.com}}}
\date{}
\maketitle

\begin{abstract}
We study physical properties of constant curvature black holes (CCBHs) in
Einstein anti-de Sitter (AdS) gravity. These objects, which are locally AdS
throughout the space, are constructed from identifications of global AdS
spacetime, in a similar fashion as Banados-Teitelboim-Zanelli (BTZ) black
hole in three dimensions. We find that, in dimensions equal or greater than
four, CCBHs have zero mass and angular momentum. Only in odd dimensions we
are able to associate a nonvanishing conserved quantity to these solutions,
which corresponds to the vacuum (Casimir) energy of the spacetime.
\end{abstract}

\section{Introduction}

The interest in three-dimensional gravity resides on the fact that black
holes in this theory are able to capture essential features of the
higher-dimensional counterparts.

It was more than two decades ago that Ba\~{n}ados, Teitelboim and
Zanelli (BTZ) found a black hole solution in $3D$ anti-de Sitter (AdS) gravity
\cite{BTZ}, characterized by mass and angular momentum, and with thermal
properties analog to rotating black holes in $D\geq 4$.

However, BTZ black hole possesses constant curvature and, therefore, it is
locally indistinguishable from global AdS space. It is only when the
solution is obtained by identifications along isometries, that one can
understand that the global structure of the spacetime is modified \cite{BHTZ}%
. In particular, mass and angular momentum appear in the holonomies computed
for a flat connection of the AdS group \cite{Holonomies}. They are also
obtained as conserved quantities coming from surface integrals in a number
of methods \cite{Carlip2+1}.

The striking properties of BTZ black holes led to some authors to look for
higher-dimensional generalizations, e.g., in the form of Constant-Curvature
Black Holes (CCBHs) \cite{Banados}.

In this paper, we study properties of CCBHs as regards mass and angular
momentum, using background-independent definitions of conserved quantities
for asymptotically AdS spacetimes.

\section{Constant Curvature Black Holes in Einstein AdS gravity}

We shall consider a pure gravity theory described by General Relativity in $%
D $ dimensions, which is given in terms of Einstein-Hilbert action%
\begin{equation}
I_{EH}=\frac{1}{16\pi G}\int\limits_{M}d^{D}x\sqrt{-g}(R-2\Lambda ).
\label{IEH}
\end{equation}%
The dynamic field is the metric $g_{\mu \nu }$, the cosmological constant is
$\Lambda $ and $R$ is the Ricci scalar, which comes from the double
contraction in the indices of the Riemann tensor $R_{\,\,\,\mu \beta \nu
}^{\alpha }=\partial _{\beta }\Gamma _{\nu \mu }^{\alpha }-\partial _{\nu
}\Gamma _{\beta \mu }^{\alpha }+\Gamma _{\beta \gamma }^{\alpha }\Gamma
_{\nu \mu }^{\gamma }-\Gamma _{\nu \gamma }^{\alpha }\Gamma _{\beta \mu
}^{\gamma }$. Because we are interested in the case of negative cosmological
constant, $\Lambda $ is given by the expression $\Lambda =-\frac{(D-1)\left(
D-2\right) }{2\ell ^{2}}$ in terms of the AdS radius $\ell $.

Arbitrary variations of the action (\ref{IEH}) with respect to the metric
give rise to the Einstein equations plus a surface term%
\begin{equation}
\delta I_{EH}=\frac{1}{16\pi G}\int\limits_{M}d^{D}x\sqrt{-g}\mathcal{E}%
_{\nu }^{\mu }\,\left( g^{-1}\delta g\right) _{\mu }^{\nu
}+\int\limits_{M}d^{D}x\,\partial _{\mu }\Theta ^{\mu },  \label{deltaIEH}
\end{equation}%
where $\mathcal{E}_{\nu }^{\mu }$ stands for the Einstein tensor%
\begin{equation}
\mathcal{E}_{\nu }^{\mu }=R_{\nu }^{\mu }-\frac{1}{2}R\,\delta _{\nu }^{\mu
}+\Lambda \delta _{\nu }^{\mu }\,,
\end{equation}%
and $\Theta ^{\mu }$ is a surface term that depends on the variation of the
Christoffel symbol.

Casting the Einstein tensor in a more convenient form%
\begin{equation}
\mathcal{E}_{\nu }^{\mu }=-\frac{1}{4}\delta _{\lbrack \nu \sigma \lambda
]}^{[\mu \alpha \beta ]}\left( R_{\alpha \beta }^{\sigma \lambda }+\frac{1}{%
\ell ^{2}}\delta _{\lbrack \alpha \beta ]}^{[\sigma \lambda ]}\right) ,
\end{equation}%
it is obvious that if the constant-curvature condition
\begin{equation}
F_{\alpha \beta }^{\sigma \lambda }=R_{\alpha \beta }^{\sigma \lambda }+%
\frac{1}{\ell ^{2}}\delta _{\lbrack \alpha \beta ]}^{[\sigma \lambda ]}=0\,,
\label{CCcondition}
\end{equation}%
is satisfied everywhere, the spacetime is a solution to AdS gravity.

In a Riemannian gravity theory in $D$ dimensions, the two-form $F$ is the
only nonvanishing part of the curvature associated to the AdS group $%
SO(D-1,2)$.

While in three-dimensional AdS gravity, a global condition $F=0$ is
equivalent to the equation of motion, imposing Eq.(\ref{CCcondition}) in
higher dimensions proves to be far more restrictive, as we shall see
below.\bigskip

\subsection{Construction of CCBHs}

In what follows, we consider solutions of Einstein equations with negative
cosmological constant which are constant-curvature black holes. This type of
solution was originally constructed by Ba\~{n}ados in Ref.\cite{Banados}.
Here, we briefly review this construction.

Let us consider a $D$-dimensional AdS space as a hypersurface defined in $%
(D+1)$-dimensional pseudo-Euclidean spacetime, subjected to the constraint
\begin{equation}
-x_{0}^{2}+x_{1}^{2}+\cdots +x_{D-2}^{2}+x_{D-1}^{2}-x_{D}^{2}=-\ell ^{2}.
\label{AdSinpE}
\end{equation}%
In particular, this surface defined by the above relation possesses a Killing vector with the components $%
\xi ^{\alpha }=(r_{+}/\ell )(0,\ldots ,0,x_{D},x_{D-1})$, which is a boost
in the $(x_{D},x_{D-1})$ plane, with a norm $\xi ^{2}=(r_{+}^{2}/\ell
^{2})(-x_{D-1}^{2}+x_{D}^{2})$.
Substituting $\xi ^{2}$ into Eq.(\ref{AdSinpE}) defines a $D-1$-dimensional hypersurface in AdS space (\ref{AdSinpE})
\begin{equation}
x_{0}^{2}=x_{1}^{2}+\cdots +x_{D-2}^{2}+\ell ^{2}\left( 1-\xi
^{2}/r_{+}^{2}\right) .  \label{AdSinXi}
\end{equation}%
In the case that $\xi=r_{+}$, the formula (\ref{AdSinXi}) leads to a null surface given by the relation
\begin{equation}
-x_{0}^{2}+x_{1}^{2}+\cdots +x_{D-2}^{2}=0.
\label{hypercone}
\end{equation}%

In order to construct a CCBH, one must identify points along the orbits of $%
\xi ^{\alpha }$. In the region $\xi ^{2}<0$ orbits of $\xi ^{\alpha }$ are timelike. However,
after the identification, they will become closed. This means that the
region $\xi ^{2}<0$ is not physical after the identification, and its
boundary $\xi ^{2}=0$ is singular in this sense. Thus spacetime of the
surface is divided into three regions: $I:=r_{+}^{2}<\xi ^{2}<\infty $, $%
II:=0<\xi ^{2}<r_{+}^{2}$ and $III:=-\infty <\xi ^{2}\leq 0$.

In order to write down explicitly the identification along the orbits of $%
\xi ^{\alpha }$, it is useful to introduce local coordinates of AdS space in
the region $\xi ^{2}>0$:
\begin{eqnarray}
x_{\tilde{\alpha}} &=&\frac{2y_{\tilde{\alpha}}}{1-y^{2}},\qquad \qquad
\qquad \tilde{\alpha}=0,\ldots ,D-2  \notag \\
x_{D-1} &=&\frac{\ell r}{r_{+}}\sinh \left( \frac{r_{+}\phi }{\ell }\right) ,
\notag \\
x_{D} &=&\frac{\ell r}{r_{+}}\cosh \left( \frac{r_{+}\phi }{\ell }\right)
\label{XgotoY}
\end{eqnarray}%
where
\begin{eqnarray}
r &=&r_{+}{(1+y^{2})}/{(1-y^{2})},  \notag \\
y^{2} &=&\eta _{\tilde{\alpha}\tilde{\beta}}y^{\tilde{\alpha}}y^{\tilde{\beta%
}}\,,  \notag \\
\eta _{\tilde{\alpha}\tilde{\beta}} &=&\mathrm{diag}(-1,1,\ldots ,1)\,.
\end{eqnarray}%
Because of the fact that the range of  coordinates is  $-\infty < x^{\alpha} <\infty $, the new variables  $-\infty <\phi <\infty $ and $-\infty
<y^{\tilde{\alpha} }<\infty $ with $-1<y^{2}<1$. Then, the metric element for the
surface (\ref{AdSinpE}) acquires the Kruskal form:
\begin{equation}
ds^{2}=\frac{\ell ^{2}(r+r_{+})^{2}}{r_{+}^{2}}\eta _{\tilde{\alpha} \tilde{\beta}
}dy^{\tilde{\alpha} }dy^{\tilde{\beta} }+r^{2}d\phi ^{2}.  \label{Kruskalmetric}
\end{equation}%
In these coordinates, the boost Killing vector has only one component $\xi
^{\phi }=1$ and its norm is $\xi ^{2}=r^{2}$. Identifying $\phi \sim \phi
+2\pi n$ one obtains a compact gravitational object that is globally of
constant curvature and the hypercone (\ref{hypercone}) becomes the horizon. Let us demonstrate that the new object is a BH. It turns
out that one can construct a Schwarzschild-like metric in the outer $I$
region. For this one needs to introduce local
\emph{spherical} coordinates
\begin{eqnarray}
y_{0} &=&f\cos \theta _{1}\sinh \frac{r_{+}t}{\ell },  \notag \\
y_{1} &=&f\cos \theta _{1}\cosh \frac{r_{+}t}{\ell },  \notag \\
y_{2} &=&f\sin \theta _{1}\sin \theta _{2}\ldots \sin \theta _{D-4}\sin
\theta _{D-3},  \notag \\
y_{3} &=&f\sin \theta _{1}\sin \theta _{2}\ldots \sin \theta _{D-4}\cos
\theta _{D-3},  \notag \\
y_{4} &=&f\sin \theta _{1}\sin \theta _{2}\ldots \cos \theta _{D-4},  \notag
\\
\ldots &=&\ldots ,  \notag \\
y_{{}_{D-3}} &=&f\sin \theta _{1}\sin \theta _{2}\cos \theta _{3},  \notag \\
y_{{}_{D-2}} &=&f\sin \theta _{1}\cos \theta _{2}  \label{Ygotospherical}
\end{eqnarray}%
where $f(r)=[{(r-r_{+})}/{(r+r_{+})}]^{1/2}$. The new coordinates are
defined for the space where $r>r_{+}$ and in the limits: $0<\theta
_{1},\theta _{2},\ldots ,\theta _{D-4}<\pi $, $0\leq \theta _{D-3}<2\pi $
and $-\infty <t<\infty $. However, the space in whole for $r>r_{+}$ is not
covered by the new coordinates because one can see from (\ref{Ygotospherical}%
) that $-1<y_{2},y_{3},\ldots ,y_{{}_{D-2}}<1$ only that is not the same as
in (\ref{XgotoY}). In the new coordinates the metric (\ref{Kruskalmetric})
acquires the Schwarzschild-like form:
\begin{equation}
ds^{2}=\ell ^{2}n^{2}(r)\left( -\cos ^{2}\theta _{1}dt^{2}+\frac{\ell ^{2}}{%
r_{+}^{2}}d\Omega _{D-3}^{2}\right) +n^{-2}(r)dr^{2}+r^{2}d\phi ^{2} \label{Smetric}
\end{equation}%
where $n^{2}(r)=(r^{2}-r_{+}^{2})/\ell ^{2}$ and, in a more explicit form,
the line element of the $(D-3)$-dim sphere can be written as
\begin{equation*}
d\Omega^{2} _{D-3}=d\theta _{1}^{2}+\sin ^{2}\theta _{1}d\Omega _{D-4}^{2}\,.
\end{equation*}%

It has been claimed that the geometry described by Eq.(\ref{Smetric})
represents a black hole with an event horizon at $r=r_{+}$. However, some of
the properties of such a black hole are rather unusual with respect to the
ones of the standard Schwarzschild solution. Indeed, the topology of CCBHs
is $R^{D-1}\times S^{1}$ instead of $R^{2}\times S^{D-2}$ (or locally flat
or hyperbolic transversal sections, as in the case of topological black
holes in AdS gravity). On the other hand, the horizon of a CCBH is
degenerated into a one-dimensional circle, whereas the horizon in
Schwarzschild black hole is a $(D-2)$-dimensional lightlike closed surface.

Furthermore, the asymptotic form ($r\rightarrow \infty $) of the metric (\ref%
{Smetric}) is not well-defined in the limit of vanishing $r_{+}$. This is a
reflection of the fact that is not possible to set $r_{+}=0$ in the
construction of CCBHs sketched in this section.

In what follows, we study physical properties of CCBHs in order to relate
the solution parameters to conserved quantities, i.e., mass and angular
momentum. Due to the fact that the boundary of the region defined by $%
r>r_{+} $ has a topology $S^{D-3}\times S^{1}$, which cannot be matched with
the asymptotic region of global AdS, no clear background can be associated
to this solution. This argument prevents the use of any
background-substraction technique \cite{background,Grishchuk-Petrov-Popova,KBL,Chen-Nester,Petrov-review} in order to define the energy of a CCBH.

\section{Background-independent charges in AdS gravity}\label{sec3}

In the previous section, we discussed on the fact that global AdS spacetime
cannot be recovered from the metric (\ref{Smetric}) by simply switching off
the parameter $r_{+}$. Therefore, we will resort to background--independent
formulas for conserved quantities in AdS gravity to evaluate the mass and
angular momentum of CCBHs.

\subsection{ Four dimensions}

In any gravity theory whose dynamics is described only by the metric field,
the variation with respect to $g_{\mu \nu }$ will give rise to the equations
of motion. Then, the energy momentum tensor of the system will be
identically zero, unless we identify it with the one coming from the matter
Lagrangian.

The alternative to introducing a matter source in the bulk is to consider a boundary stress tensor,
idea that was developed some time ago by Brown and York in Ref.\cite{Brown-York}%
. In particular, in a Gauss-normal coordinate frame%
\begin{equation}
ds^{2}=N^{2}(r)dr^{2}+h_{ij}(x,r)dx^{i}dx^{j}  \label{Gauss-normal}
\end{equation}%
this\emph{\ quasilocal} stress tensor is obtained as the variation respect
to the metric $h_{ij}$, that is,%
\begin{equation*}
T^{ij}=\frac{2}{\sqrt{-h}}\frac{\delta I}{\delta h_{ij}},
\end{equation*}%
assuming that the field equations hold in the bulk.

In the above formula, $I$ stands for Einstein-Hilbert action supplemented by
suitable boundary terms, such that it is stationary under arbitrary
variations of the boundary metric.

As it has been extensively discussed in the literature, the proper way of
getting rid of normal derivatives in the boundary metric in the surface term
is adding a Gibbons-Hawking boundary term%
\begin{equation}
I=I_{EH}-\frac{1}{8\pi G}\int\limits_{\partial M}d^{D-1}x\sqrt{-h}K\,.
\end{equation}%
The boundary $\partial M$ is defined in terms of the foliation (\ref{Gauss-normal})
as located at constant $r$.
The term at the boundary is proportional to the trace of the extrinsic curvature, which is
defined as%
\begin{equation}
K_{ij}=-\frac{1}{2N}\partial _{r}h_{ij}  \label{extrinsicK}
\end{equation}%
in the coordinate frame (\ref{Gauss-normal}).

The action $I$ produces a quasilocal stress tensor which coincides with the
definition of canonical momentum $\pi ^{ij}$ for the radial foliation of the
spacetime (\ref{Gauss-normal}), that is,

\begin{equation}
\pi ^{ij}=\frac{1}{8\pi G}\left( K^{ij}-h^{ij}K\right) \,.  \label{Piij}
\end{equation}%
Because the line element (\ref{Gauss-normal}) describes the spacetime as an
infinite series of concentric cylinders, the conservation of (\ref{Piij})
for a given $r$ is a consequence of Einstein equations in that frame.

When used  for General Relativity with zero cosmological constant, this stress tensor provides a sensible definition of conserved quantities
as surface integrals in the asymptotic region \cite{Brown-York}. However, in asymptotically AdS gravity
this boundary energy-momentum tensor produces charges which are divergent even for
three-dimensional AdS black holes.

In the context of anti-de Sitter gravity/Conformal Field Theory (AdS/CFT)
correspondence, there is a systematic way to construct finite conserved
charges for asymptotically AdS spacetimes. In order to have a finite
variation of the action, one needs to add a \emph{counterterm }series $%
\mathcal{L}_{ct}(h,\mathcal{R},\nabla \mathcal{R})$, with dependence on the
boundary curvature $\mathcal{R=R}(h)$ and covariant derivatives of it, such
that the total action is renormalized%
\begin{equation}
I_{ren}=I+\int\limits_{\partial M}d^{D-1}x\,\mathcal{L}_{ct}(h,\mathcal{R}%
,\nabla \mathcal{R}).
\end{equation}%
With the addition of local counterterms, the Brown-York stress tensor adopts
the general form%
\begin{equation}
T^{ij}=\pi ^{ij}+\frac{2}{\sqrt{-h}}\frac{\delta \mathcal{L}_{ct}}{\delta
h_{ij}}\,.
\end{equation}

In particular, in $D=4$, the counterterms required for a proper
regularization of AdS gravity action are \cite{Balasubramanian-Kraus,EJM}%
\begin{equation}
\mathcal{L}_{ct}=\frac{\sqrt{-h}}{8\pi G}\left( \frac{2}{\ell }+\frac{\ell }{%
2}\,\mathcal{R}(h)\right) \,,
\end{equation}%
such that the quasilocal stress tensor takes the form%
\begin{equation}
T_{i}^{j}=\frac{1}{8\pi G}\left( K_{i}^{j}-\delta _{i}^{j}K+\frac{2}{\ell }%
\delta _{i}^{j}-\ell \left( \mathcal{R}_{i}^{j}(h)-\frac{1}{2}\delta _{i}^{j}%
\mathcal{R}(h)\right) \right) .  \label{Tij BK}
\end{equation}%
Any asymptotically AdS spacetime can be described by a metric that has a
divergence or order two in the radial coordinate, as one approches the
asymptotic region. In particular, holographic techniques employ a
Fefferman-Graham frame \cite{FG} to realize the Asymptotically Locally AdS
condition, i.e., the fact that the curvature tends to a constant at the
boundary. In particular, this asymptotic behavior implies that the extrinsic
properties of the boundary, given by $K_{ij}$ can be expressed as a series
of intrinsic quantities
\begin{equation}
K_{i}^{j}=\frac{1}{\ell }\delta _{i}^{j}+\ell S_{i}^{j}(h)+\mathcal{O(R}^{2}%
\mathcal{)}
\end{equation}%
where%
\begin{equation*}
S_{i}^{j}(h)=\mathcal{R}_{i}^{j}(h)-\frac{1}{4}%
\delta _{i}^{j}\mathcal{R}(h) \,,
\end{equation*}%
is the Schouten tensor of the boundary metric $h_{ij}$. It is not difficult
to see that the linear terms in the extrinsic curvature can be regarded as a
truncation of an expression that is quadratic in $K$, because the quasilocal
stress tensor can be re-written as%
\begin{equation*}
T_{i}^{j}=\frac{\ell }{16\pi G}\,\delta _{\left[ inp\right] }^{\left[ jkl%
\right] }\left( \frac{1}{2}\mathcal{R}_{kl}^{np}(h)-K_{k}^{n}K_{l}^{p}+\frac{%
1}{\ell ^{2}}\delta _{k}^{p}\delta _{l}^{n}\right) \,,
\end{equation*}%
up to $\mathcal{O(R}^{2}\mathcal{)}$ terms.
Anti-symmetrization of the indices leaves the last formula in the form%
\begin{eqnarray}
T_{i}^{j} &=&\frac{\ell }{32\pi G}\,\delta _{\left[ inp\right] }^{\left[ jkl%
\right] }\left( \mathcal{R}%
_{kl}^{np}(h)-K_{k}^{n}K_{l}^{p}+K_{k}^{p}K_{l}^{n}+\frac{1}{\ell ^{2}}%
\delta _{\lbrack kl]}^{[pn]}\right) ,  \notag \\
&=&\frac{\ell }{32\pi G}\,\delta _{\left[ inp\right] }^{\left[ jkl\right]
}\left( R_{kl}^{np}+\frac{1}{\ell ^{2}}\delta _{\lbrack kl]}^{[pn]}\right) ,
\label{TAdScurv}
\end{eqnarray}%
where we have used the Gauss-Codazzi relation (\ref{Codazzi1}). The quantity
in brackets corresponds to the boundary components of the AdS curvature,
which is the tensor that appears in the left hand side of Eq.(\ref%
{CCcondition}), and that is identically vanishing for global AdS spacetime.
Thus, $T_{i}^{j}$ also vanish identically for a spacetime with constant
curvature everywhere.

On the other hand, for Einstein gravity, where the Ricci tensor is $R_{\mu
\nu }=-\frac{3}{\ell ^{2}}g_{\mu \nu }$, the Weyl tensor
\begin{equation}
W_{\mu \nu }^{\alpha \beta }=R_{\mu \nu }^{\alpha \beta }-\frac{1}{2}%
R_{[\mu }^{[\alpha }\delta _{\nu ]}^{\beta ]}+\frac{1}{6}R\delta
_{\lbrack \mu \nu ]}^{[\alpha \beta ]}\,,  \label{Weyldef}
\end{equation}%
becomes%
\begin{equation}
W_{\mu \nu }^{\alpha \beta }=F_{\mu \nu }^{\alpha \beta }\,,  \label{WisF}
\end{equation}%
when Einstein equations hold.

That means that the quasilocal stress tensor is nothing but a projection of
the Weyl tensor%
\begin{equation}
T_{i}^{j}=\frac{\ell }{32\pi G}\,\delta _{\left[ inp\right] }^{\left[ jkl%
\right] }W_{kl}^{np}=-\frac{\ell }{8\pi G}\,W_{jr}^{ir}\,,
\end{equation}%
where we have used the fact that single and double trace of the Weyl are
zero. In a more covariant form, the boundary stress tensor is the electric
part of the Weyl tensor
\begin{equation}
T_{i}^{j}=-\frac{\ell }{8\pi G}\,W_{j\nu }^{i\beta }\,n^{\nu }n_{\beta
}\,=E_{i}^{j},
\end{equation}%
where $n$ is a normal vector to the boundary. This implies that the
conformal mass definition for AAdS spaces provided by Ashtekar and Magnon in
Ref.\cite{Ashtekar-Magnon} also gives a vanishing mass for CCBHs.

The electric part of the Weyl tensor in Eq.(\ref{TAdScurv}) is, in turn, a
truncation (up to quadratic order in the boundary curvature) of a charge
density obtained from the addition of a topological invariant
(Gauss-Bonnet), which is cubic in the extrinsic curvature \cite%
{Olea2n,Miskovic-Olea4D}%
\begin{equation*}
T_{i}^{j}=\frac{\ell }{32\pi G}\,\delta _{\left[ inp\right] }^{\left[ mkl%
\right] }K_{m}^{j}\left( \mathcal{R}%
_{kl}^{np}(h)-K_{k}^{n}K_{l}^{p}+K_{k}^{p}K_{l}^{n}+\frac{1}{\ell ^{2}}%
\delta _{\lbrack kl]}^{[pn]}\right) \,.
\end{equation*}%
We stress the fact that the comparison is possible by using the standard
asymptotic behavior of the metric and the curvature for AAdS spacetimes.
Additional terms appearing in the different formulas for energy in AdS
gravity may play a role in a modified asymptotic behavior or for a boundary
located at a finite cutoff.

But what is of relevance here is the fact that the equivalence between
different notions of conserved quantities shows that CCBHs always have zero
mass and angular momentum.

\subsection{General even-dimensional case $(D=2n)$}

For the purpose of computation of the conserved quantities for the solution (%
\ref{Smetric}) we will employ the charges derived within Kounterterm
regularization scheme for AdS gravity \cite{Olea2n,OleaKTs}.

The conservation of Noether charges, calculated as surface integrals, is a
consequence of the existence of a conserved current $J^{\mu }$. Indeed, the
quantity $Q[\xi ]=\int_{\partial M}\sqrt{-h}n_{\mu }\,J^{\mu }$ is a
constant of motion, where $h$ is the determinant of the metric of the
hypersurface orthogonal to the normal \ vector $n_{\mu }$. In terms of the
radial foliation (\ref{Gauss-normal}), this quantity gives rise to the energy
and angular momentum (and, in principle, other conserved quantities) enclosed
by the sphere at that radius.

For the line element on $\partial M$, we take the set of coordinates\textbf{%
\ }$x^{i}=\left( t,x^{m}\right) $, such that it adopts an ADM form%
\begin{equation}
h_{ij}\,dx^{i}dx^{j}=-\tilde{N}^{2}(t)dt^{2}+\sigma _{mn}(dx^{m}+\tilde{N}%
^{m}dt)(dx^{n}+\tilde{N}^{n}dt)\,,\qquad \sqrt{-h}=\tilde{N}\sqrt{\sigma }\,,
\label{ADMboundary}
\end{equation}%
which is generated by the timelike unit vector $u_{i}=(-\tilde{N},\vec{0})$.
The metric $\sigma _{mn}$ represents the geometry of the subspace $\Sigma
_{\infty }$, which is the spatial section of the asymptotic region (at
constant time).

Whenever the radial component of the current can be globally expressed on
the boundary as a derivative of the
\begin{equation}
\sqrt{-g}\,J^{r}=\partial _{j}\left[ \sqrt{-h}\,\xi ^{i}\,\left(
q_{i}^{j}+q_{(0)i}^{j}\,\right) \right] ,
\end{equation}%
the Noether theorem provides the conserved charges $Q[\xi ]$ of the theory%
\begin{equation}
Q[\xi ]=q\left[\xi \right] +q_{(0)}\left[\xi \right] \,,  \label{totalQ}
\end{equation}%
where each term is expressed as surface integrals on $\Sigma _{\infty }$ as
\begin{eqnarray}
q\left[ \xi \right] &=&\int\limits_{\Sigma _{\infty }}d^{D-2}x\,\sqrt{\sigma
}\,u_{j}\,\xi ^{i}\,q_{i}^{j}\,,  \label{mass} \\
q_{(0)}\left[ \xi \right] &=&\int\limits_{\Sigma _{\infty }}d^{D-2}x\,\sqrt{%
\sigma }\,u_{j}\,\xi ^{i}\,q_{(0)i}^{j}\,.  \label{vacuum
energy}
\end{eqnarray}%
for a given set of asymptotic Killing vectors $\{\xi \}$. The splitting of
the charge density in two parts is motivated by the fact that $q_{(0)i}^{j}$
gives rise to the vacuum energy in odd-dimensional AAdS spacetimes. In turn,
in even dimensions, it vanishes identically.

In the Kounterterm method, both $q_{i}^{j}$ and $q_{(0)i}^{j}$ are given as
polynomials of the intrinsic and extrinsic curvatures.

The charge density in the $2n$-dimensional case is%
\begin{equation}
q_{i}^{j}=\frac{1}{16\pi G}\frac{\left( -1\right) ^{n}\ell ^{2n-2}}{%
2^{n-2}\left( 2n-2\right) !}\,\delta _{\lbrack i_{1}i_{2}\cdots
i_{2n-1}]}^{[jj_{2}\cdots j_{2n-1}]}\,K_{i}^{i_{1}}\left( {R}%
_{j_{2}j_{3}}^{i_{2}i_{3}}\cdots {R}_{j_{2n-2}j_{2n-1}}^{i_{2n-2}i_{2n-1}}-%
\frac{\left( -1\right) ^{n-1}}{\ell ^{2n-2}}\,\delta _{\lbrack
j_{2}j_{3}]}^{[i_{2}i_{3}]}\cdots \delta _{\lbrack
j_{2n-2}j_{2n-1}]}^{[i_{2n-2}i_{2n-1}]}\right) \,.  \label{qij even}
\end{equation}%
It is possible to prove that the CCBH metric, for fixed $r$ and $\phi$ and after a proper time rescaling,
represents a $(D-2)-$ dimensional de Sitter space. This implies the absence of globally defined timelike Killing vector in this geometry \cite{Holst-Peldan}.
Employing a global coordinate chart to describe CCBHs leads to a time-dependent line element \cite{Cai}.

The alternative to the use of Killing vectors to define conserved quantities in a time-dependent geometry is the definition of Kodama vectors given in Ref.\cite{Kodama}. However, Kodama's construction provides an answer for simple metrics, but it cannot be straightforwardly extended to an arbitrary gravitational object.

However, the two terms under the bracket in the formula (\ref{qij even}) can be always factorized
by the AdS curvature%
\begin{equation}
q_{i}^{j}=\frac{1}{16\pi G}\frac{\left( -1\right) ^{n}\ell ^{2n-2}}{%
2^{n-2}\left( 2n-2\right) !}\,\delta _{\lbrack i_{1}i_{2}\cdots
i_{2n-1}]}^{[jj_{2}\cdots j_{2n-1}]}\,K_{i}^{i_{1}}\left(
R_{j_{2}j_{3}}^{i_{2}i_{3}}+\frac{1}{\ell ^{2}}\delta _{\lbrack
j_{2}j_{3}]}^{[i_{2}i_{3}]}\right) P_{j_{4}\cdots j_{2n-1}}^{i_{4}\cdots
i_{2n-1}}({R,\delta })\,,
\label{qijevenP}
\end{equation}%
where $P(R,\delta )$ is a polynomial of the spacetime Riemann tensor and the
antisymmetric delta of rank $2$, which is more conveniently written in a
parametric integral form%
\begin{equation}
P_{j_{4}\cdots j_{2n-1}}^{i_{4}\cdots i_{2n-1}}=(n-1)\int\limits_{0}^{1}dt
\left[ \left( 1-t\right) \,R_{j_{4}j_{5}}^{i_{4}i_{5}}-\frac{t}{\ell ^{2}}%
\,\delta _{\lbrack j_{4}j_{5}]}^{[i_{4}i_{5}]}\right] \cdots \left[ \left(
1-t\right) \,R_{j_{2n-2}j_{2n-1}}^{i_{2n-2}i_{2n-1}}-\frac{t}{\ell ^{2}}%
\,\delta _{\lbrack j_{2n-2}j_{2n-1}]}^{[i_{2n-2}i_{2n-1}]}\right] \,.
\end{equation}

This fact implies that the integrand of the charge is identically zero for any spacetime
which is globally of constant curvature. Thus, for CCBHs (\ref{Smetric}) the
expression (\ref{qijevenP}) gives zero identically, what is manifest even
before evaluating the explicit metric.

\subsection{General odd-dimensional case $(D=2n+1)$}

In odd dimensions, the expressions for the charge density $q_{i}^{j}$ and $%
q_{(0)i}^{j}$ are given by%
\begin{eqnarray}
q_{i}^{j} &=&-\frac{1}{2^{n-2}}\,\delta _{\lbrack ki_{1}\cdots
i_{2n-1}]}^{[jj_{1}\cdots j_{2n-1}]}\,K_{i}^{k}\delta _{j_{1}}^{i_{1}}\left[
\frac{1}{16\pi G\,(2n-1)!}\,\delta _{\lbrack
j_{2}j_{3}]}^{[i_{2}i_{3}]}\cdots \delta _{\lbrack
j_{2n-2}j_{2n-1}]}^{[i_{2n-2}i_{2n-1}]}\right.  \notag \\
&&\qquad \qquad +\left. nc_{2n}\int\limits_{0}^{1}dt\,\left( {R}%
_{j_{2}j_{3}}^{i_{2}i_{3}}+\frac{t^{2}}{\ell ^{2}}\,\delta _{\lbrack
j_{2}j_{3}]}^{[i_{2}i_{3}]}\right) \cdots \left( {R}%
_{j_{2n-2}j_{2n-1}}^{i_{2n-2}i_{2n-1}}+\frac{t^{2}}{\ell ^{2}}\,\delta
_{\lbrack j_{2n-2}j_{2n-1}]}^{[i_{2n-2}i_{2n-1}]}\right) \right] \,,
\label{qij_odd}
\end{eqnarray}%
where $c_{2n}$ is a constant, given by%
\begin{equation}
c_{2n}=\frac{1}{16\pi G}\frac{\left( -1\right) ^{n}\ell ^{2n-2}}{%
2^{2n-2}n\left( n-1\right) !^{2}}\,.  \label{c2n}
\end{equation}%
The expression for the charge can be factorized as%
\begin{equation}
q_{i}^{j}=-\frac{nc_{2n}}{2^{n-1}}\,\delta _{\lbrack ki_{1}\cdots
i_{2n-1}]}^{[jj_{1}\cdots j_{2n-1}]}\,K_{i}^{k}\delta _{j_{1}}^{i_{1}}\left(
{R}_{j_{2}j_{3}}^{i_{2}i_{3}}+\frac{1}{\ell ^{2}}\,\delta _{\lbrack
j_{2}j_{3}]}^{[i_{2}i_{3}]}\right) \tilde{P}_{j_{4}\cdots
j_{2n-1}}^{i_{4}\cdots i_{2n-1}}({R,\delta })\,,  \label{qij odd}
\end{equation}%
where $\tilde{P}_{j_{4}\cdots j_{2n-1}}^{i_{4}\cdots i_{2n-1}}({R,\delta })$
is a Lovelock-type polynomial of degree $(n-2)$ in the Riemann tensor, which
takes the following form when written in terms of a double parametric
integral
\begin{align}
\tilde{P}_{j_{4}\cdots j_{2n-1}}^{i_{4}\cdots i_{2n-1}}=2\left( n-1\right)
\,& \int\limits_{0}^{1}dt\int\limits_{0}^{1}ds\,\left[ s\left(
R_{j_{4}j_{5}}^{i_{4}i_{5}}+\frac{1}{\ell ^{2}}\,\delta _{\lbrack
j_{4}j_{5}]}^{[i_{4}i_{5}]}\right) +\frac{t^{2}-1}{\ell ^{2}}\,\delta
_{\lbrack j_{4}j_{5}]}^{[i_{4}i_{5}]}\right] \times \cdots  \notag \\
& \cdots \times \left[ s\left( R_{j_{2n-2}j_{2n-1}}^{i_{2n-2}i_{2n-1}}+\frac{%
1}{\ell ^{2}}\,\delta _{\lbrack
j_{2n-2}j_{2n-1}]}^{[i_{2n-2}i_{2n-1}]}\right) +\frac{t^{2}-1}{\ell ^{2}}%
\,\delta _{\lbrack j_{2n-2}j_{2n-1}]}^{[i_{2n-2}i_{2n-1}]}\right] .
\label{Ptilde}
\end{align}

Because $q_{i}^{j}$ is factorized by $\left( {R}_{j_{2}j_{3}}^{i_{2}i_{3}}+%
\frac{1}{\ell ^{2}}\,\delta _{\lbrack j_{2}j_{3}]}^{[i_{2}i_{3}]}\right) $,
then it is identically zero for a spacetime of constant curvature everywhere, as it is the case for CCBHs.

The part (\ref{vacuum energy}) of the total charge density, which does not vanish
for globally constant-curvature spacetimes, has an explicit expression%
\begin{equation}
q_{(0)j}^i  = -\frac{nc_{2n}}{2^{n-2}}\delta _{\lbrack ki_{1}\cdots
i_{2n-1}]}^{[jj_{1}\cdots j_{2n-1}]}\,\left( K_{i}^{k}\,\delta _{j_{1}}^{i_{1}}+K_{j_{1}}^{k}\delta
_{i}^{i_{1}}\right) \int\limits_{0}^{1}du\,u\,\mathcal{F}%
_{j_{2}j_{3}}^{i_{2}i_{3}}(u)\times \cdots \times \mathcal{F}%
_{j_{2n-2}j_{2n-1}}^{i_{2n-2}i_{2n-1}}(u)\,.  \label{q0general}
\end{equation}
where%
\begin{equation}
\mathcal{F}_{kl}^{ij}(u)=\mathcal{R}_{kl}^{ij}-u^{2}\left(
K_{k}^{i}K_{l}^{j}-K_{l}^{i}K_{k}^{j}\right) +\frac{u^{2}}{\ell ^{2}}%
\,\delta _{\left[ kl\right] }^{\left[ ij\right] }  \label{Fcalu}
\end{equation}%

Due to the lack of globally-defined timelike Killing vectors, we compute the integrand $q_{(0)t}^t$, in order to compare with the results obtained for topological AdS black holes given in the Appendix
\ref{E0Topological}. This quantity contains information on certain holographic modes of AAdS spaces which give rise to a non-zero vacuum energy. The main advantage is the fact that the expression for $q_{(0)t}^t$ exists in any odd dimension, what allows us to perform a generic computation for $D=2n+1$.

As a warmup computation, let us first consider the five-dimensional case. The
extrinsic and intrinsic curvatures expressions for CCBHs are given (see Appendix \ref{staticCCBH}).
Once we set the components $i$ and $j$ as $t$ in Eq.(\ref{q0general}), we obtain
\begin{eqnarray}
q_{(0)t}^t  &=&-\frac{\ell ^{2}}{64\pi G}%
\delta _{\lbrack
n_{1}n_{2}n_{3}]}^{[m_{1}m_{2}m_{3}]}\,\left(
K_{t}^{t}\,\delta _{m_{1}}^{n_{1}}-K_{m_{1}}^{n_{1}}\delta _{t}^{t}\right)
\times   \notag \\
&&\times \int\limits_{0}^{1}du\,u\,\left[ \mathcal{R}%
_{m_{2}m_{3}}^{n_{2}n_{3}}-u^{2}\left(
K_{m_{2}}^{n_{2}}K_{m_{3}}^{n_{3}}-K_{m_{3}}^{n_{2}}K_{m_{2}}^{n_{3}}\right)
+\frac{u^{2}}{\ell ^{2}}\,\delta _{\lbrack m_{2}m_{3}]}^{[n_{2}n_{3}]}\right]
\,.  \notag
\end{eqnarray}%
Due to the transversal symmetries of CCBHs, the indices $m,n$ can be
separated into the ones corresponding to the sphere $S^{2}$ and an
additional azimuthal angle $\phi $.
\begin{equation}
q_{(0)_t}^t =-\frac{\ell ^{2}}{128\pi G}\,\delta
_{[ b_{1}b_{2}]}^{[ a_{1}a_{2}]}\,\left( K_{t}^{t}-K_{\phi }^{\phi }\right)\,\left[ \mathcal{R}_{a_{1}a_{2}}^{b_{1}b_{2}}-\frac{1}{2}\left(
K_{a_{1}}^{b_{1}}K_{a_{2}}^{b_{2}}-K_{a_{2}}^{b_{1}}K_{a_{1}}^{b_{2}}\right)
+\frac{1}{2\ell ^{2}}\,\delta _{\lbrack a_{1}a_{2}]}^{[b_{1}b_{2}]}\right] .
\end{equation}
Due to the fact that $\mathcal{F}_{\phi a}^{\phi b}(u)=0$,
all other contributions vanish.
Using the explicit form for the intrinsic
and extrinsic curvatures for this type of black holes, the above expression
turns into\,
\begin{eqnarray}
q_{(0)t}^t  &=&\frac{1}{64\ell\pi G}\left (\frac{r_+^4}{r^4}\right )+\mathcal{O}\left (\frac{1}{r^{6}}\right )\,,
\end{eqnarray}%
where we used the formulae in Appendix \ref{Deltas}. This simple example already reflects
an unusual dependence on the parameters $r_+$ and $\ell$ in the vacuum energy density, when compared
to the one for 5D Schwarzschild-AdS black holes (See Appendix \ref{E0Topological}).

Computing the general case, the integrand takes the form
\begin{equation}
 q^t_{(0)t} = -\frac{nc_{2n}}{2^{n-2}}\int_0^1du\,u\,\delta^{[a_1\cdots a_{2n-2}]}_{[b_1\cdots b_{2n-2}]}\left ( K_t^t-
 K_\phi^\phi \right )\mathcal{F}_{a_1a_2}^{b_1b_2}(u)\times\cdots\times \mathcal{F}_{a_{2n-3}a_{2n-2}}^{b_{2n-3}b_{2n-2}}(u)\,.
\end{equation}
The values of $\mathcal{F}$ are identical for all the $a,b$ indexes.
Summing up all the contributions
\begin{eqnarray}
 q^t_{(0)t} & = & 2nc_{2n}(2n-2)!\left ( \frac{r_+^2}{\ell r\sqrt{r^2-r_+^2}} \right )\left ( \frac{r_+^2}{\ell^2(r^2-r_+^2)}\right )^{n-1}\int_0^1du\,u(1-u^2)^{n-1}  \nonumber \\
& = & \frac{(-1)^n}{8\pi G\ell (2n-1)}\frac{r_+^{2n}}{r(r^2-r_+^2)^{n-\frac{1}{2}}}\frac{(2n-1)!!^2}{(2n)!} \,.
\end{eqnarray}
The leading order for the previous expression is
\begin{equation}\label{q0tt}
q^t_{(0)t}=\frac{(-1)^n}{8\pi G\ell (2n-1)}\frac{r_+^{2n}}{r^{2n}}\frac{(2n-1)!!^2}{(2n)!}\,+\mathcal{O}\left (\frac{1}{r^{2n+2}}\right )\,.
\end{equation}

A similar computation was carried for the five-dimensional CCBH by Cai in Ref. \cite{Cai}, using a quasilocal stress tensor, properly renormalized by the addition of local counterterms to the AdS gravity action. In this case, the metric is written in a manifestly time-dependent form, such that the boundary energy also exhibits a dependence on the parameter $r_{+}$. It is also claimed in that reference that proper rescalings of the coordinates in CCBH solution can remove the dependence of $r_{+}$ in the boundary stress tensor. What is left, however, contains powers of the AdS radius $\ell$ over $G$ such that the result cannot be understood as the Casimir energy of a boundary CFT. Furthermore, the above result carries, in any odd dimension, the opposite sign with respect to formula (\ref{q0topological}) for topological AdS black holes.
The result in Eq.(\ref{q0tt}) suggests that the above conclusion is also valid in higher odd dimensions: it is not possible to attribute a well-defined zero point energy to CCBHs.

\subsection{Rotating CCBHs}

The construction of spinning CCBHs relies on the identification along
isometries which are different respect to the ones of the static case. This
procedure is equivalent to boosting the static metric along the $t-\phi $
plane \cite{Banados}
\begin{eqnarray}
t & \rightarrow & \beta t \frac{r_+}{\ell^2}+(\phi-\Omega \beta t)\frac{r_-}{\ell}\,, \notag \\
\phi & \rightarrow &  \beta t \frac{r_-}{\ell^2}+(\phi-\Omega \beta t)\frac{r_+}{\ell}\,, \label{t-phi}
\end{eqnarray}
which introduces $r_{+}$ and $r_{-}$ as two arbitrary
constants ($r_{+}>r_{-}$) in the metric. This is motivated by a similar transformation acting on the static BTZ black hole in $3D$, which
indeed generates angular momentum.
In Ref.\cite{Banados}, it is stated that a different choice of the parameters $\beta$ and $\Omega$ does not modify the conserved charges of this type of locally AdS solutions.

In a way, one could understand this statement along the line of reasoning of the previous section. Indeed, no matter what particular values $\beta$ and $\Omega$ take, the mass and angular momentum of the solution are identically zero, what is evident from Eq.(\ref{qij odd}).
In order to complete the present discussion, we can compute the vacuum charge density (\ref{q0general}) in any odd dimension. The integrand in Eq.(\ref{q0general}) for the boosted metric can be expanded as
\begin{eqnarray}
q^t_{(0)t} & = &  -\frac{nc_{2n}}{2^{n-2}}\delta^{[a_1\cdots a_{2n-2}]}_{[b_1\cdots b_{2n-2}]}\int_0^1du\,u\left[ \left ( K_t^t-K_\phi^\phi\right) \mathcal{F}_{a_1a_2}^{b_1b_2}(u)+ \right . \nonumber \\
 && \left . \left ( K_t^t\delta_{a_1}^{b_1}-K_{a_1}^{b_1}\right )\mathcal{F}_{\phi a_2}^{\phi b_2}(u)- K_t^\phi\delta_{a_1}^{b_1}\mathcal{F}_{\phi a_2}^{t b_2}(u) \right ] \mathcal{F}_{a_3a_4}^{b_3b_4}(u)\times\cdots\times \mathcal{F}_{a_{2n-3}a_{2n-2}}^{b_{2n-3}b_{2n-2}}(u)\,.
\end{eqnarray}
Summing  up all the contributions
\begin{eqnarray}
q^t_{(0)t} & = & 2nc_{2n}(2n-2)!\int_0^1du\,u(1-u^2)^{n-1}\frac{r_+^4\left (r_+^2-r_-^2+2nr_-(r_+\Omega \ell-r_-)\right  )}{r\ell^3(r_+^2-r_-^2)(r^2-r_+^2)^{\frac{3}{2}}}\left (\frac{r_+^2}{\ell^2(r^2-r_+^2)}\right )^{n-2} \nonumber \\
& = & \frac{(-1)^n}{8\pi G\ell (2n-1)}\frac{(2n-1)!!^2}{(2n)!}\frac{r_+^{2n}\left (r_+^2-r_-^2+2nr_-(r_+\Omega \ell-r_-)\right  )}{r(r_+^2-r_-^2)(r^2-r_+^2)^{n-\frac{1}{2}}}\,.
\end{eqnarray}
The leading order for this expression is
\begin{equation} \label{qttr}
q^t_{(0)t} = \frac{(-1)^n}{8\pi G\ell (2n-1)}\frac{(2n-1)!!^2}{(2n)!}\frac{r_+^{2n}\left (r_+^2-r_-^2+2nr_-(r_+\Omega \ell-r_-)\right  )}{r^{2n}(r_+^2-r_-^2)}+\mathcal{O}\left (\frac{1}{r^{2n+2}}\right )
\end{equation}
In Ref.\cite{Banados-Gomberoff-Martinez} a particular choice of the parameters $\Omega=\frac{r_-}{r_+\ell}$ and $\beta=r_+$ is taken in order to define a five-dimensional Lorentzian  rotational CCBH. Plugging in these values on Eq.(\ref{qttr}), we obtain
\begin{equation}
q^t_{(0)t} = \frac{(-1)^n}{8\pi G\ell (2n-1)}\frac{(2n-1)!!^2}{(2n)!}\frac{r_+^{2n}}{r^{2n}}+\mathcal{O}\left (\frac{1}{r^{2n+2}}\right )\,,
\end{equation}
what is the same result as in the static case.\\
Therefore, the presence of terms of the type $dtd\phi $ in the line element does not bring in rotation in the solution.
That means that the fall-off of these crossed terms is such that they do not contribute to surface integrals defined at radial
infinity.\\
In sum, the boost (\ref{t-phi}) does not generate a new solution, as it is unable to change
the physical parameters of CCBHs.

\section{Discussion and conclusions}

We have shown that, in even-dimensional AdS gravity, mass and angular
momentum for CCBHs is always zero. This is a consequence of the fact that
the conserved charge formulas for AdS gravity can be always factorized by
the AdS curvature or, equivalently, the Weyl tensor.

This result is not surprising in the light of a recent work \cite{KJMO
Conformal}, where it is shown that, for AAdS spaces, conformal mass
definition is linked to the addition of Kounterterms. It is then reassuring
the fact that two energy definitions in AdS gravity give the same zero
result for CCBH mass.

In principle, one could always go against this reasoning claiming that the mass for a
black hole may also be obtained from the integration of matter stress tensor at
the right hand side of the Einstein equation. However, delta-type
singularities are hidden inside the horizon and not needed by
methods that compute conserved charges as surface integrals in the asymptotic region.

But what looks undeniable is the fact that conserved charges computed within
Kounterterm regularization provide the correct answer for a large class of
AAdS solutions regardless a particular form of the matter energy-momentum
tensor. This is confirmed by the relation between Kounterterm charges and
conformal mass definition in every dimension \cite{Ashtekar-Magnon,Ashtekar-Das,KJMO Conformal}.

In odd dimensions, a nonvanishing value for $q^{t}_{(0)t}$ does not make the
situation more promising for CCBHs. Indeed, the formulas that usually give
the mass and angular momentum for AAdS solutions also vanish identically in
this case. For static CCBHs, the vacuum energy contains a rather unusual
dependence on the parameter $r_{+}$. Even though one can eliminate this dependence by
coordinate rescalings, the result is not proportional to $\ell^{2n-1}/G$, what is the only sensible
value one can interpret as a Casimir energy for the CFT on the boundary of AdS gravity in $D=2n+1$
dimensions.

In the rotating case, the second parameter in the solution, $r_{-}$, does
not appear anywhere in the conserved quantities, nor in the expression for
the vacuum energy. This result contrasts with the interpretation of $r_{-}$ as related to angular momentum
in the original reference \cite{Banados}, because it is a well-known fact that, in $%
D=2n+1$ dimensions, the vacuum energy for Kerr-AdS black holes, depends on
the rotation parameters $\{a_{i}\}_{i=1..n}$ \cite{CVJ,Das-Mann,Pa-Sk,Ol-Ol}.
Even in the simplest case (Myers-Perry-AdS), with a single rotation
parameter $a$, the vacuum energy depends on $r_{-}$.

If one attempts to consider CCBHs as solutions of Einstein-Gauss-Bonnet or a
more general Lovelock gravity theory with AdS asymptotics, the main
conclusions of this paper would remain the same. In fact, for EGB AdS
gravity, Kounterterm regularization will give rise to conserved charges
whose formulas are factorizable by the corresponding AdS curvature \cite%
{Kofinas-Olea EGB, KJMO Conformal EGB},%
\begin{equation}
\tilde{F}_{\mu \nu }^{\alpha \beta }=R_{\mu \nu }^{\alpha \beta }+\frac{1}{%
\ell _{\mathrm{eff}}^{2}}\delta _{\lbrack \mu \nu ]}^{[\alpha \beta ]}\,,
\end{equation}%
in terms of an effective AdS radius%
\begin{equation}
\frac{1}{\ell _{\mathrm{eff}}^{2}}=\frac{1}{2(D-3)(D-4)\alpha }\left( 1\pm
\sqrt{1-\frac{4(D-3)(D-4)\alpha }{\ell ^{2}}}\right) \,,  \label{eq:leffpm}
\end{equation}%
where $\alpha $ is the Gauss-Bonnet coupling.

A black hole which is globally of constant curvature can indeed have mass in
three-dimensional AdS gravity, as it is the case of BTZ solution \cite%
{BTZ,BHTZ}. This is understood in the light of the general formulas for the
conserved charges in odd dimensions Eqs.(\ref{qij_odd}) and (\ref{q0general}). In
three dimensions, there are no enough indices in Eq.(\ref{qij_odd}) to
produce an expression proportional to the AdS curvature (or, equivalently,
the Weyl tensor), as it happens in higher dimensions. As a consequence, the
equivalence to formula (\ref{qij_odd}) vanishes identically in 3D. Then, it
is Eq.(\ref{q0general}) the formula responsible for the mass and angular momentum
for BTZ black hole \cite{MOR}.

As AdS gravity in 3D is derived from a Chern-Simons action for $SO(2,2)$
group, the previous result can be extended to higher odd dimensions. Indeed,
in Lovelock-Chern-Simons gravity, it is a formula proportional to Eq.(\ref%
{q0general}) the one that produces the mass of black holes in that theory \cite%
{Miskovic-Olea DCG}. As a matter of fact, this is the only case one can
atribute a nonvanishing mass to CCBHs \cite{CCBH in CS}.

An independent approach that might shed some light on the problem of vanishing charges for
CCBHs is the computation of holonomies, motivated by the result in $3D$ AdS gravity \cite{Holonomies}.
However, this computation would require the existence of non-contractible curves which enclose the \emph{horizon} in higher dimensions.
On the other hand, in the context of supersymmetry, arbitrary values of the parameters in the CCBH solution should produce an obstruction
to the existence of globally-defined Killing spinors \cite{Steif,Aros-Romo}.

\bigskip

\section{Acknowledgments}

We would like to thank O. Miskovic for insightful discussions. P.G. is a UNAB M.Sc. Scholarship Holder. This work was
funded in part by FONDECYT Grants No. 1090357 and 1131075, UNAB Grant DI-1336-16/R and CONICYT Grant DPI 20140115.

\section*{Appendices}
\appendix

\section{Kronecker delta of rank $p$ \ \label{Deltas}}

The totally-antisymmetric Kronecker delta of rank $p$ is defined as the
determinant
\begin{equation}
\delta _{\left[ \mu _{1}\cdots \mu _{p}\right] }^{\left[ \nu _{1}\cdots \nu
_{p}\right] }:=\left\vert
\begin{array}{cccc}
\delta _{\mu _{1}}^{\nu _{1}} & \delta _{\mu _{1}}^{\nu _{2}} & \cdots &
\delta _{\mu _{1}}^{\nu _{p}} \\
\delta _{\mu _{2}}^{\nu _{1}} & \delta _{\mu _{2}}^{\nu _{2}} &  & \delta
_{\mu _{2}}^{\nu _{p}} \\
\vdots &  & \ddots &  \\
\delta _{\mu _{p}}^{\nu _{1}} & \delta _{\mu _{p}}^{\nu _{2}} & \cdots &
\delta _{\mu _{p}}^{\nu _{p}}%
\end{array}%
\right\vert \,.
\end{equation}%
A contraction of $k\leq p$ indices in the Kronecker delta of rank $p$
produces a delta of rank $p-k$,
\begin{equation}
\delta _{\left[ \mu _{1}\cdots \mu _{k}\cdots \mu _{p}\right] }^{\left[ \nu
_{1}\cdots \nu _{k}\cdots \nu _{p}\right] }\,\delta _{\nu _{1}}^{\mu
_{1}}\cdots \delta _{\nu _{k}}^{\mu _{k}}=\frac{\left( N-p+k\right) !}{%
\left( N-p\right) !}\,\delta _{\left[ \mu _{k+1}\cdots \mu _{p}\right] }^{%
\left[ \nu _{k+1}\cdots \nu _{p}\right] }\,,
\end{equation}%
where $N$ is the range of indices.

\section{Extrinsic and intrinsic curvatures for static CCBHs \ \label%
{staticCCBH}}

The radial foliation of the spacetime
\begin{equation}
ds^{2}=N^{2}(r)dr^{2}+h_{ij}(x,r)dx^{i}dx^{j}
\end{equation}%
implies the Gauss-Codazzi relations for the spacetime Riemann tensor%
\begin{eqnarray}
{R}_{kl}^{ir} &=&\frac{1}{N}\,\left( \nabla _{l}K_{k}^{i}-\nabla
_{k}K_{l}^{i}\right) \,,  \label{Codazzi2} \\
{R}_{kr}^{ir} &=&\frac{1}{N}\,\left( K_{k}^{i}\right) ^{\prime
}-K_{l}^{i}\,K_{k}^{l}\,,  \label{Codazzi3} \\
{R}_{kl}^{ij} &=&\mathcal{R}_{kl}^{ij}(h)-K_{k}^{i}\,K_{l}^{j}+K_{l}^{i}%
\,K_{k}^{j}\,,  \label{Codazzi1}
\end{eqnarray}%
where $\nabla _{l}=\nabla _{l}(\Gamma _{ij}^{k})$ is the covariant
derivative defined in the Christoffel symbol of the boundary ${\Gamma }%
_{ij}^{k}(g)=\Gamma _{ij}^{k}(h)$.

On the other hand, the boundary metric for CCBHs takes the block-diagonal
form%
\begin{equation}
h_{ij}=\left(
\begin{array}{ccc}
-(r^{2}-r_{+}^{2})\cos ^{2}\theta _{1} & 0 & 0 \\
0 & (r^{2}-r_{+}^{2})\frac{\ell ^{2}}{r_{+}^{2}}\gamma _{ab} & 0 \\
0 & 0 & r^{2}%
\end{array}%
\right)
\end{equation}%
where the boundary indices split as $i=(t,a,\phi )$ and the metric $\gamma
_{ab}$ is the metric of the unit sphere $S^{D-3}.$\bigskip

From the line element on the boundary metric%
\begin{equation}
h_{ij}\,dx^{i}dx^{j}=(r^{2}-r_{+}^{2})\left( -\cos ^{2}\theta _{1}dt^{2}+%
\frac{\ell ^{2}}{r_{+}^{2}}d\Omega _{D-3}\right) +r^{2}d\phi ^{2}
\end{equation}%
it is easy to read off the lapse function and the induced metric in an ADM
foliation with vanishing shift functions

\begin{eqnarray}
h_{ij}\,dx^{i}dx^{j} &=&-\tilde{N}^{2}dt^{2}+\sigma _{mn}dx^{m}dx^{n}
\notag \\
&=&-\tilde{N}^{2}dt^{2}+r^{2}\left( \gamma _{ab}dy^{a}dy^{b}+d\phi
^{2}\right) \,,
\end{eqnarray}%
such that%
\begin{equation}
\tilde{N}^{2}=(r^{2}-r_{+}^{2})\cos ^{2}\theta _{1}  \label{Ntilde}
\end{equation}%
where $y^{a}$ and $\gamma _{ab}$ (with $a,b=\{1,...,D-3\}$) are the angles
and the metric of $(D-3)$-dimensional sphere, respectively, and $\phi $ is
an additional azimuthal angle. The lapse functions sets the only novanishing
component of the timelike normal vector as $u_{t}=-(r^{2}-r_{+}^{2})^{1/2}|%
\cos \theta _{1}|$, and the determinant $\sigma $ of the spatial metric is
given by%
\begin{equation}
\sqrt{\sigma }=r^{D-2}\left( \frac{\ell }{r_{+}}\right) ^{D-3}\left( 1-\frac{%
r_{+}^{2}}{r^{2}}\right) ^{\frac{D-3}{2}}\sqrt{\gamma _{D-3}}\text{%
\thinspace .}  \label{det sigma}
\end{equation}

The components of the extrinsic curvature for the metric (\ref{Smetric}) are
\begin{eqnarray}
K_{t}^{t} &=&-\frac{r}{\ell }\frac{1}{(r^{2}-r_{+}^{2})^{1/2}}\,,  \notag \\
K_{b}^{a} &=&-\frac{r}{\ell }\frac{1}{(r^{2}-r_{+}^{2})^{1/2}}\delta
_{b}^{a}\,,  \notag \\
\qquad K_{\phi }^{\phi } &=&-\frac{(r^{2}-r_{+}^{2})^{1/2}}{r\ell }\,.
\label{extrinsicKCCBH}
\end{eqnarray}%
Recalling the fact that the solution (\ref{Smetric}) is a constant-curvature
spacetime, the intrinsic curvature $\mathcal{R}_{kl}^{ij}$ can be directly
obtained by Gauss-Codazzi relations. In doing so, the only non-vanishing
components of the boundary Riemann tensor are%
\begin{eqnarray}
\mathcal{R}_{tb}^{ta}(h) &=&\frac{r_{+}^{2}}{\ell ^{2}}\,\frac{1}{%
r^{2}-r_{+}^{2}}\,\delta _{b}^{a}\,,  \notag \\
\mathcal{R}_{cd}^{ab}(h) &=&\frac{r_{+}^{2}}{\ell ^{2}}\,\frac{1}{%
r^{2}-r_{+}^{2}}\,\delta _{\lbrack cd]}^{[ab]}\,\,.  \label{curvatureRCCBH}
\end{eqnarray}
\noindent
In the case of rotating CCBH, the components of the extrinsic curvatures are
\begin{eqnarray}
K_t^t & = & -\frac{r^2(r_+^2-r_-^2)+r_+^2r_-(r_--r_+\Omega \ell)}{\ell r (r_+^2-r_-^2)\sqrt{r^2-r_+^2}}\,, \\
K_\phi^\phi & = & -\frac{r^2(r_+^2-r_-^2)+r_+^3(\Omega\ell r_--r_+)}{\ell r (r_+^2-r_-^2)\sqrt{r^2-r_+^2}}\,, \\
K_t^\phi & = & -\frac{r_+^2\beta (\Omega\ell (r_+^2-r_-^2)-r_+r_-(1+\Omega^2 \ell^2) )}{\ell^2 r (r_+^2-r_-^2)\sqrt{r^2-r_+^2}}\,, \\
K_\phi^t & = & -\frac{r_+^3r_-}{ r\beta (r_+^2-r_-^2)\sqrt{r^2-r_+^2}}\,, \\
K_a^b & = & -\frac{r}{\ell\sqrt{r^2-r_+^2}}\delta^a_b\,.
\end{eqnarray}
\noindent
The boundary Riemman tensors
\begin{eqnarray}
\mathcal{R}_{ab}^{cd} & = & \frac{r_+^2}{\ell^2 (r_+^2-r_-^2)\sqrt{r^2-r_+^2}}\delta_{[ab]}^{[cd]}\,, \\
\mathcal{R}_{\phi a}^{\phi b} & = & \frac{r_+^2r_-(\Omega \ell r_+-r_-)}{\ell^2(r_+^2-r_-^2)(r^2-r_+^2)}\delta_a^b\,, \\
\mathcal{R}_{ta}^{\phi b} & = & \frac{r_+^3r_-}{\beta\ell(r_+^2-r_-^2)(r^2-r_+^2)}\delta_a^b\,.
\end{eqnarray}
And the values of $\mathcal{F}_{\alpha\beta}^{\mu\nu}(u)$
\noindent
\begin{eqnarray}
\mathcal{F}^{a_1a_2}_{b_1b_2}(u) & = & (1-u^2)\frac{r_+^2}{\ell^2 (r_+^2-r_-^2)\sqrt{r^2-r_+^2}}\delta^{[a_1a_2]}_{[b_1b_2]}\,, \\
\mathcal{F}^{\phi a}_{\phi b}(u) & = & (1-u^2)\frac{r_+^2r_-(\Omega \ell r_+-r_-)}{\ell^2(r_+^2-r_-^2)(r^2-r_+^2)}\delta^a_b\,, \\
\mathcal{F}_{t b}^{\phi a}(u) & = & (1-u^2)\frac{r_+^3r_-}{\beta\ell(r_+^2-r_-^2)(r^2-r_+^2)}\delta^a_b\,.
\end{eqnarray}

\section{Vacuum energy in Topological Black Holes in AdS Gravity} \label{E0Topological}

In order to compare to the results in Section \ref{sec3}, we here review the
computation of the zero-point energy for globally AdS spacetime.

Static topological black holes are solutions of the Einstein equations with
negative cosmological constant, which are described by the metric%
\begin{equation}
ds^{2}=-f^{2}(r)dt^{2}+f^{-2}(r)dr^{2}+r^{2}d\Sigma _{D-2}^{2}
\label{staticBH}
\end{equation}%
where the line element of the transversal section $\Sigma ^{(k)}$ is%
\begin{equation}
d\Sigma _{D-2}^{2}=\gamma _{mn}^{(k)}\,dy^{m}dy^{n}\,,
\end{equation}%
and%
\begin{equation}
f^{2}(r)=k-\frac{2G\mu }{r^{D-3}}+\frac{r^{2}}{\ell ^{2}}\,,  \label{f2r}
\end{equation}%
where the parameter $\mu $ is the mass density. The topological parameter $%
k=+1,0,-1$ denotes the curvature of the transversal section, which can be a
sphere, a locally flat surface or a hyperboloid, respectively.

Using the general formula for the vacuum energy for AAdS spacetimes (\ref%
{q0general}) for the particular choice of the timelike vector $u_{j}$ and a timelike Killing vector $\xi ^{i}$,
leads to the expression
\begin{eqnarray}
q_{(0)t}^t & = & \frac{nc_{2n}}{2^{n-2}}\delta _{[ m_{1}\cdots
m_{2n-1}]}^{[n_{1}\cdots n_{2n-1}]}\,\left( K_{t}^{t}\,\delta _{n_{1}}^{m_{1}}-K_{n_{1}}^{m_{1}}\delta
_{t}^{t}\right)  \notag \\
&& \int\limits_{0}^{1}du\,u\,\mathcal{F}%
_{n_{2}n_{3}}^{m_{2}m_{3}}(u)\times \cdots \times \mathcal{F}%
_{n_{2n-2}n_{2n-1}}^{m_{2n-2}m_{2n-1}}(u)\,.
\end{eqnarray}%
The components of the tensorial quantities defined at the boundary --which
are relevant for the evaluation of the above formula-- are given by
\begin{equation}
K_{t}^{t}\,=-f^{\prime }(r),\qquad K_{n_{1}}^{m_{1}}=-\frac{f(r)}{r}\,\delta
_{n_{1}}^{m_{1}}
\end{equation}%
for the extrinsic curvature and
\begin{equation}
\mathcal{F}_{n_{2}n_{3}}^{m_{2}m_{3}}(u)=\frac{1}{r^{2}}\left[ k-u^{2}\left(
f^{2}(r)-\frac{r^{2}}{\ell ^{2}}\right) \right] \delta _{\lbrack
n_{2}n_{3}]}^{[m_{2}m_{3}]}\,,
\end{equation}%
for the parametric curvature (\ref{Fcalu}).

In the static black hole ansatz (\ref{staticBH}), the vacuum energy integrand adopts
the form
\begin{eqnarray}
q_{(0)t}^t &=&2nc_{2n}(2n-1)!\,\left( f^{\prime}(r)-\frac{f(r)}{r}\right)\left ( \frac{1}{r^{2n-2}} \right )
\int\limits_{0}^{1}du\,u\,\left[ k-u^{2}\left( f^{2}(r)-\frac{r^{2}}{\ell ^{2}}%
\right) \right] ^{n-1},
\end{eqnarray}%
which, taking the value of the metric function in Eq.(\ref{f2r}),
adopts a much simpler form
\begin{eqnarray}
q_{(0)t}^t &=&2nc_{2n}(2n-1)!\,\left(-\frac{\ell k}{r^2}+\mathcal{O}(r^{1-D})\right )\left ( \frac{1}{r^{2n-2}} \right ) \notag \\
&& \int\limits_{0}^{1}du\,u\,\left[ k-u^{2}\left( k-\frac{2G\mu }{r^{D-3}}\right) \right]^{n-1},
\end{eqnarray}
in the limit $r\rightarrow \infty$ for the leading order
\begin{equation}
q_{(0)t}^t=-c_{2n}(2n)!\frac{ k^{n}\ell}{r^{2n}}\int\limits_{0}^{1}du\,u\,\left( 1-u^{2}\right) ^{n-1}+\cdots.
\end{equation}%
When performed the trivial integration in the parameter $u$, the standard result
of the vacuum energy integrand for topological
AdS black holes is recovered
\begin{equation} \label{q0topological}
q_{(0)t}^t = \frac{(-1)^{n-1}\ell^{2n-1}}{8\pi G}\frac{(2n-1)!!^2}{(2n)!}\frac{k^n}{r^{2n}}\,. \notag \\
\end{equation}
The standard result of the vacuum energy for topological
AdS black holes is recovered when integrated in the transversal section
considering $\sqrt{\sigma}=r^{2n-1}\sqrt{\gamma^{(k)}}$
\begin{eqnarray}
E_{0}&=&\int\limits_{\Sigma _{\infty }}d^{2n-1}x\,\left( -f(r)\right) q_{(0)t}^t\sqrt{\sigma} \notag \\
&=& \frac{\ell ^{2n-2}}{8\pi G}(-k)^{n}\,\frac{(2n-1)!!^{2}}{\left( 2n\right)
!}Vol\left( \Sigma ^{(k)}\right)
\end{eqnarray}
 where $\gamma^{(k)}$ is the determinant of the surface $\Sigma ^{(k)}$.
%%%%%%%%%%%%%%%%%%%%%%%%%%%%%%%%%%%%%%%%%%%%%%%%%%%%%%%%%%%%%%%%%%%%%%%%%%%%%%%

\end{document}